\documentclass[pra,reprint,amsmath,amssymb,superscriptaddress]{revtex4-1}
\usepackage{graphicx}
\usepackage{mathrsfs}
\usepackage{color}
\usepackage{url}

\newcommand{\G}{\Gamma}

\begin{document}

% Use the \preprint command to place your local institutional report number 
% on the title page in preprint mode.
% Multiple \preprint commands are allowed.
%\preprint{}

\title{Quantitative Atomic Spectroscopy for Primary Thermometry} %Title of paper

% repeat the \author .. \affiliation  etc. as needed
% \email, \thanks, \homepage, \altaffiliation all apply to the current author.
% Explanatory text should go in the []'s, 
% actual e-mail address or url should go in the {}'s for \email and \homepage.
% Please use the appropriate macro for the type of information

% \affiliation command applies to all authors since the last \affiliation command. 
% The \affiliation command should follow the other information.

\author{Gar-Wing Truong}
\email[]{Gar-Wing.Truong@physics.uwa.edu.au}
%\homepage[]{Your web page}
%\thanks{}
%\altaffiliation{}
\affiliation{Frequency Standards and Metrology Research Group, School of Physics, The University of Western Australia, Perth, WA 6009, Australia}

\author{Eric F. May}
%\email[]{}
%\homepage[]{Your web page}
%\thanks{}
%\altaffiliation{}
\affiliation{Centre for Energy, School of Mechanical and Chemical Engineering, The University of Western Australia, Perth, WA 6009, Australia}

\author{Thomas M. Stace}
%\email[]{}
%\homepage[]{Your web page}
%\thanks{}
%\altaffiliation{}
\affiliation{School of Mathematics and Physics, University of
Queensland, Brisbane, QLD 4072, Australia}

\author{Andr\'{e} N. Luiten}
%\email[]{}
%\homepage[]{Your web page}
%\thanks{}
%\altaffiliation{}
\affiliation{Frequency Standards and Metrology Research Group, School of Physics, The University of Western Australia, Perth, WA 6009, Australia}

% Collaboration name, if desired (requires use of superscriptaddress option in \documentclass). 
% \noaffiliation is required (may also be used with the \author command).
%\collaboration{}
%\noaffiliation

\date{\today}

\begin{abstract}
Quantitative spectroscopy has been used to measure accurately the Doppler-broadening of atomic transitions in $^{85}$Rb vapor. By using a conventional platinum resistance thermometer and the Doppler thermometry technique, we were able to determine $k_B$ with a relative uncertainty of $4.1\times 10^{-4}$, and with a deviation of $2.7\times 10^{-4}$ from the expected value. Our experiment, using an effusive vapour, departs significantly from other Doppler-broadened thermometry (DBT) techniques, which rely on weakly absorbing molecules in a diffusive regime. In these circumstances, very different systematic effects such as magnetic sensitivity and
optical pumping are dominant. Using the model developed recently by Stace and Luiten, we estimate the perturbation due to optical pumping of the measured $k_B$ value was less than $4\times 10^{-6}$. The effects of optical pumping on atomic and molecular DBT experiments is mapped over a wide range of beam size and saturation intensity, indicating possible avenues for improvement. We also compare the line-broadening mechanisms, windows of operation and detection limits of some recent DBT experiments.
\end{abstract}

% insert suggested PACS numbers in braces on next line

\maketitle %\maketitle must follow title, authors, abstract and \pacs

% Body of paper goes here. Use proper sectioning commands. 
% References should be done using the  \cite, \ref, and \label commands
\section{Introduction}
%\label{}
%\subsection{}
%\subsubsection{}
An ability to measure absolute thermodynamic temperature is currently confined to an exclusive list of national standards laboratories. Such measurements require primary thermometry expertise and highly specialised equipment, making it impractical for wider-scale use. Instead, for practical purposes, temperature scales such as ITS-90 \cite{prestonthomas1990}, are used as an approximation to the true thermodynamic temperature, even though they are known to have relative uncertainties at the $10^{-4}$ level \cite{pitre}. Accordingly, there is now a global push to develop more convenient methods of primary thermometry that could make thermodynamic temperature measurements more broadly available \cite{CCT_05_31}. A second motivating factor driving recent renewed interest in primary thermometry is the call by the Bureau des Internationale de Poids et Measures' to remeasure the Boltzmann constant, $k_B$, using a wide range of techniques in preparation for the redefinition of the kelvin in 2011 \cite{fellmuth2006,mills2006}.

At present, the recommended value for $k_B$ is derived primarily from one acoustic measurement performed by Moldover \textit{et. al.} \cite{moldover}. It would be preferable if $k_B$ could be remeasured using different methods with comparable uncertainties so that any systematic errors can be identified. Recently, Bord\'{e} suggested that the known\cite{demtroeder1981} temperature dependence of the spectral linewidth of absorption features, previously exploited for measurments of stellar atmospheric temperatures\cite{unsold2001}, can be used as a new technique for accurate primary thermometry\cite{borde}. This Doppler-broadening thermometry (DBT) method has since been experimentally realized and while currently less precise than other primary methods such as acoustic or dielectric constant gas thermometry, it suffers from very different types of systematic errors \cite{fischer2005}. In contrast to those techniques which measure the macroscopic quantity $RT$, the DBT method can directly probe the microscopic thermal energy $k_B T$ by measuring the characteristic spectral width of molecular/atomic transitions, thereby avoiding the uncertainties in the Avogadro constant $N_A$. 

The first DBT measurements performed by Daussy \textit{et. al.} and Casa \textit{et. al.} have used molecular gases at 1-130 Pa and determined $k_B$ with a relative uncertainty \cite{daussy2007,casa2008,castrillo2009} of order $10^{-4}$. More recent work has improved the statistical uncertainty \cite{djerroud2009} to $4\times 10^{-5}$. In this paper, we present results from an atomic Rubidium (Rb) DBT system at $3\times 10^{-5}$ Pa. The use of an alkali metal atom absorber presents some distinct spectroscopic differences that allow us to explore new experimental regimes within the DBT method. In doing so, we have identified some advantages in using an atomic system and encountered new challenges whose resolutions demand a deep understanding and detailed unification of light-matter interactions and gas dynamical theories \cite{stace2010}. Whilst this work has been motivated by primary thermometry, we believe that it has wider implications in all fields requiring high-resolution or precision, quantitative spectroscopy. 

The primary difference between a low-pressure atomic vapor system, like Rb, and molecular experiments is that atomic motion is effusive, so collisions are extremely rare. An advantage of moving to this effusive regime is the avoidance of pressure-induced systematic changes, such as collisional line-shape perturbations \cite{wende1981}. This removes the need to extrapolate results to an equivalent zero-pressure value. Further, by using a sealed vapor-pressure reference gas cell, we avoid any unintended pressure drifts associated with a more sophisticated variable-pressure gas chamber. In this study, we chose Rb which has a high optical cross section, ensuring that the absorption signal-to-noise is satisfactory in spite of the very low vapor density. This choice also gives the ability to operate in a range where high quality lasers are readily available and where silicon detectors, with their extremely good linearity  \cite{Shin2005} and quantum efficiency, can be used. At room temperature, it was possible to use a smaller cell than those in the molecular experiments for the same absorption depth, which was convenient for thermal control. 

On the other hand, there are also disadvantages to using atomic vapors. Care is required to ensure low levels of residual magnetic fields which can lead to the broadening of spectral lines by lifting the degeneracy of Zeeman levels. In the transition  used in this experiment ($D2$ line of $^{85}$Rb), optical pumping between unresolved hyperfine transitions contributed the greatest potential for systematic error by perturbing the lineshape. In addition, the comparability of the natural lifetime, beam transit time and the Rabi period in our situation leads to optical pumping effects that requires either an accurate model of the gas kinetics \cite{stace2010} or the use of a low power probe beam to avoid lineshape perturbations. We followed the second route here and show here that the resulting perturbation is at the 4 parts-per-million (ppm) level. We note, however, that it should be possible to operate at higher power levels and use the model of Stace and Luiten \cite{stace2010} to correct the results.

The remainder of this paper is organised as follows. In Sec. II, we describe the experimental setup and data acquisition for a Rb DBT experiment. The theoretical absorption coefficient lineshape for unperturbed atoms and the derivation of a suitable lineshape profile for extracting the quantity $k_BT$ from the acquired spectra are then described in Sec. III.  Our experimental results are presented in Sec. IV, followed by a discussion about systematic uncertainties in Sec. V. In Sec. VI, we discuss the relative merits and challenges of the atomic approach in more detail. We will also discuss how the use of atomic vapors might be extended to a wider temperature range.

\section{Method and Experiment}\label{sec:exp}
 \begin{figure}
 \includegraphics[scale=1]{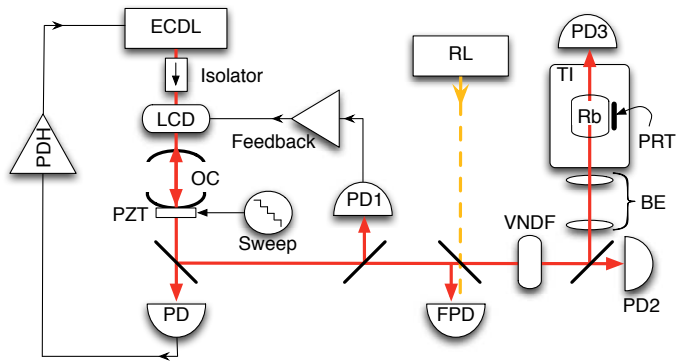}%
 \caption{(Color Online) A schematic diagram of the optical circuit and electronic control systems. ECDL: extended cavity diode laser; LCD: variable optical depth liquid crystal display; OC: optical cavity; PZT: annular piezo-electric stack; PDH: Pound-Drever-Hall top-of-resonance feedback control loop; RL: reference laser; VNDF: variable neutral density filter; BE: beam expander lenses; PRT: ITS-90 calibrated platinum resistance thermometer; TI: thermal isolator; (F)PD: (fast) photodetector. \label{fig:setup}}%
 \end{figure}

 \begin{figure}
 \includegraphics[scale=1]{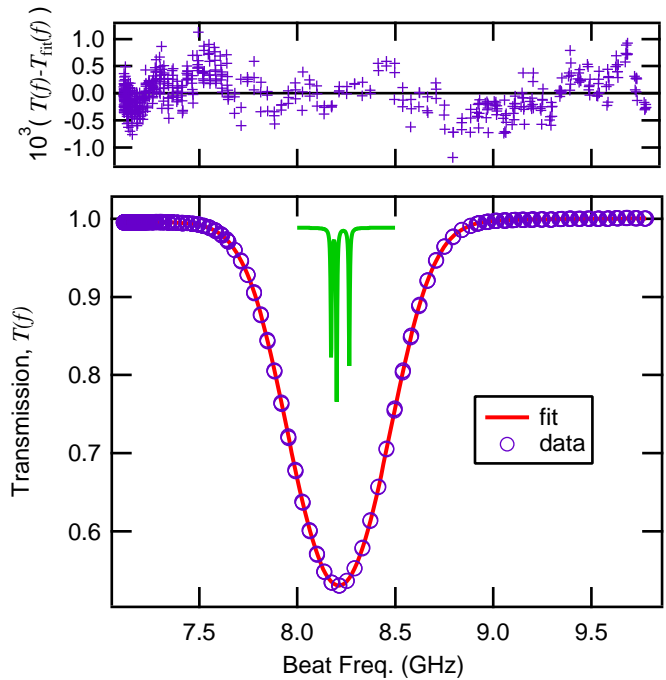}%
 \caption{(Color online) The lower panel shows a typical spectrum (blue circles) obtained from Rb vapour at 295 K ($\sim$550 points) and the fit (red) using Eqn. (\ref{eqn:fitfunction}). The Lorentzians (green) indicate both the location of the underlying hyperfine transitions and their relative strengths (in arbitrary units). The upper panel displays the residuals to the fit.\label{graph:typical}}%
 \end{figure}

A schematic of the experimental setup used in this work is shown in Fig. \ref{fig:setup}. The $k_BT$ product was determined from spectroscopic measurements of unresolved hyperfine transitions (from the $5^2S_{1/2}$, $F=2 \rightarrow 5^2P_{3/2}$, $F=1,2,3$ states) in the $D2$ line of $^{85}$Rb at 780 nm. A custom-built extended cavity diode laser (ECDL) was locked to a tunable optical Fabry-P\'{e}rot cavity (OC) using the Pound-Drever-Hall (PDH) technique \cite{black2001}, which suppressed acoustic and other fluctuations of the laser frequency. We tuned the laser's wavelength by changing the length of the cavity. A sweep generator (Sweep) was used to displace one of the OC cavity mirrors by driving a piezo-electric stack actuator (PZT). A stepwise sweep was used to reduce the asynchronicity between measurements of the temperature, frequency and absorbed power. An LCD variable optical attenuator was used to keep the intensity transmitted through the cavity, which was monitored by the photodetector PD1, at a constant. This suppressed variations in the amount of optical pumping associated with power changes occurring throughout the scan. 

The optical frequency difference between the probe and a reference laser (RL) was measured using a fast photodetector (FPD) and a high bandwidth counter. The reference beam was derived from a tunable Ti:Sapphire laser locked to a temperature-controlled, ultra-low expansion optical cavity. The stability of the reference laser ($\sim$1 kHz) contributed negligible uncertainty to the optical frequency measurement. A variable neutral density filter (VNDF) allowed measurements to be made at incident probe powers below that set by the LCD intensity control system, without affecting intensity or frequency stabilities. A photodetector (PD2) located just before the absorption cell was used to correct for any residual probe power changes. A beam expander (BE) was used to give the largest beam possible through the cell (2 cm in diameter) to maximise the input power for a given intensity to achieve a high signal-to-noise ratio. The power transmitted through a 10 cm long Rb absorption cell was measured on a third photodetector (PD3). The Rb cell was encased in a passive thermal isolator (TI) whose temperature was monitored using a using a platinum resistance thermometer (PRT) calibrated to ITS-90 with 30 mK uncertainty. The maximum temperature difference across the isolator was measured to be less than 7 mK using auxiliary thermometers (not shown). The thermal relaxation time constant of the isolator ($\sim$2 hours) was much larger than the time required for a single sweep (10 mins). 

The outputs of PD2 and PD3 that monitored the incident probe and the transmitted powers, respectively, were recorded using digital multimeters with 5-digit resolution. A third multimeter was used to make four-wire resistance measurements of the PRT. A computer-based data acquisition system centrally controlled and logged the three digital multimeters and the high bandwidth counter outputs. Before each set of measurements at a fixed probe power level, the DC voltages on PD2 and PD3 caused by stray lights, were recorded in the absence of the probe beam. These backgrounds were removed from the detector readings before reconstructing the Doppler-broadened absorption spectrum by dividing the output of PD3 by that of PD2. 

\section{Extracting $k_BT$ from Spectra}\label{sec:ExtractkB}
In a low density gas cell, the absorption coefficient of an isolated spectral line is the convolution between a Lorentzian and a Gaussian function, known as a Voigt profile \cite{demtroeder1981}. The Lorentzian component can be written as $L(\nu)=(1+[(\nu-\nu_0)/\Gamma]^2)^{-1}$, where $\nu$ is the optical frequency and $\nu_0$ is the atomic transition frequency. This component arises from the finite lifetime of the upper state and has a full-width-at-half-maximum (FWHM) of $\G=1/(2\pi\tau)$, where $\tau$ is the total upper state lifetime. Various phenomena including collisional, transit time and power broadening can perturb the width of this Lorentzian. It was important in this experiment to keep these perturbations at negligible levels. 

The Gaussian component has the form of $G(\nu)=\exp(-[(\nu-\nu_0)/\Delta\nu_D]^2)$, where the $1/e$-half-width of the Gaussian component is related to the thermal energy $k_BT$ as:
\begin{eqnarray}
k_BT = \frac{mc^2}{2}\left(\frac{\Delta\nu_D}{\nu_0}\right)^2\label{eqn:dopWidth}
\end{eqnarray} 
and arises from the thermal motion of the atoms. Here $c$ is the speed of light, $m$ is the mass of the absorbing atom and $\nu_0$ is the absolute transition frequency. For Rb, the latter two parameters are known with relative uncertainties of $5\times 10^{-8}$ and $5\times 10^{-11}$, respectively \cite{bradley1999, arimondo1977}, which are negligible for this experiment. Whilst it was possible to compute the absorption coefficient $V(\nu,\Delta\nu_D)=L(\nu)\otimes G(\nu)$ by performing a convolution integral, this was a computationally inefficient procedure. Instead, a numerical routine described by Humlicek \cite{humlicek1982} implemented in the scientific data analysis program \textit{Igor Pro}\footnote{\textit{Igor Pro} ver. 5.0.4.8 Wavemetrics, Inc., Portland, OR., USA.} was used to generate the Voigt function. The relative accuracy of this algorithm was better than $3\times 10^{-5}$ and no consideration of these numerical errors was required in the following analysis.

The observed transmission spectrum was determined by Beer's Law \cite{demtroeder1981}, $T(\nu)=\exp(-V(\nu,\Delta\nu_D) L)$, where $L$ is the optical path length of the probe inside the absorbing gas. Since this experiment concerns frequency differences, it was convenient to shift the frequency origin by subtracting away the absolute optical frequency of the $^{87}$Rb $F=2\rightarrow 1$ hyperfine transition. We denote these frequency differences with $f$ to avoid confusion with the absolute frequencies $\nu$. We performed least-squares regression of the transmission spectra to Eqn. (\ref{eqn:fitfunction}) to determine $\Delta\nu_D$.
\begin{eqnarray}
T(f)\!=\!Af\!+\!B\exp\!\!\left(\!-C \sum_{i=1}^3 S_i V_i(f\!-\!f_i\!-\!f_{\textrm{cav}},\Delta\nu_D)\!\!\right)\!
\label{eqn:fitfunction}
\end{eqnarray}
Here, $T(f)$ is the sum of three, unresolved hyperfine transitions, indicated schematically in Fig. \ref{graph:typical}; the $V_{i}(f-f_{i}-f_{\textrm{cav}})$ are the Voigt functions corresponding to each hyperfine transition, each centered on frequency $f_{i}$ and with relative strength $S_{i}$; and $A$, $B$ and $C$ are three of the five parameters adjusted in the regression. The $S_{i}$ were fixed at values derived from the Clebsch-Gordan coefficients while the $f_{i}$ were fixed at values determined by Arimondo \textit{et. al.} \cite{arimondo1977} Of the adjustable parameters, $A$ and $B$ account for, respectively, a residual background and imperfect background normalisation (to a value of 1) of the spectra. $C$ is the on-resonance absorption depth determined by the number density and the cell length. The two other adjustable parameters, $f_\textrm{cav}$ and $\Delta\nu_{D}$ were contained within, and were common to the three Voigt functions, with $f_\textrm{cav}$ allowing for the arbitrary (but fixed) frequency offset of the reference laser from the lowest frequency hyperfine line. The Boltzmann constant was obtained from the Gaussian width $\Delta\nu_{D}$ using Eqn. (\ref{eqn:dopWidth}). 

In principle, it is possible to perform the regression allowing every parameter in Eqn. \ref{eqn:fitfunction} to be adjustable. However, in this case, the returned parameters from the least-squares fit show strong correlations between a number of the free parameters. This leads a larger uncertainty range in those parameters. In particular, we find strong correlations in the $S_i$, and also in $\Gamma$ and $\Delta\nu_D$ which have a correlation coefficient of \mbox{-0.96}. Operating in an intensity range where it is possible to fix the Lorentz width and the strengths $S_i$ to known values, we avoid an artificial inflation of the uncertainties associated with the fitted Doppler width due to these correlations. Our theoretical model (see Section \ref{sec:systematics}) guides us on the upper intensity.

 \begin{figure}
 \includegraphics[scale=1]{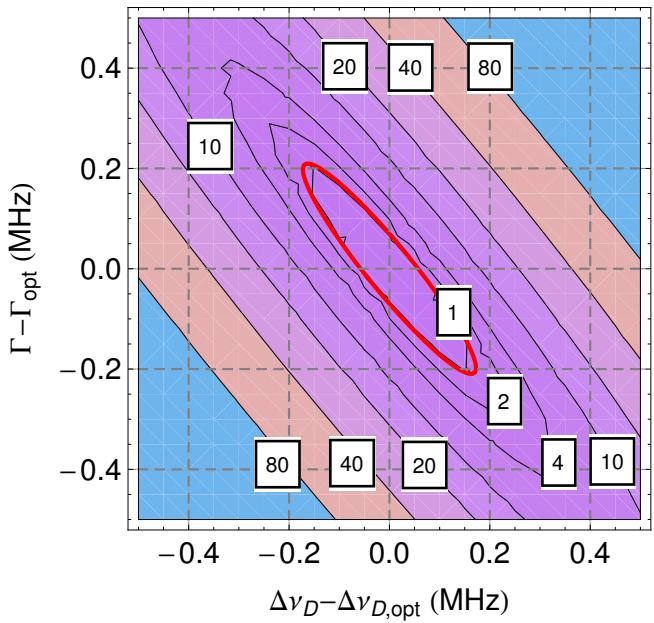}%
 \caption{(Color Online) The $(\chi^2-\chi^2_\textrm{opt})$-surface as a function of change in Lorentzian and Gaussian widths, denoted $\Gamma-\Gamma_\textrm{opt}$ and $\Delta\nu_D-\Delta\nu_{D\textrm{,opt}}$ respectively, show that there is a high degree of correlation between these fit parameters. The subscripts \textit{opt} denote the optimum quantities determined from the least-squares fit, which were 2.75 MHz and 308.0 MHz, for the Lorentz and Gaussian widths, respectively. The thicker, red contour is generated from the covariance matrix and is intended to guide the eye. \label{graph:chisquare}}%
 \end{figure}

\begin{table}
\caption{Values of the fixed constants used in Eqn. \ref{eqn:fitfunction} for the relative transition strengths, $S_i$, and their frequency offsets \cite{arimondo1977}, $f_i$. The relative transition strengths are exact numbers, whilst there is a relative uncertainty of order $1\times 10^{-5}$ associated with the differences between the hyperfine transition frequencies.\label{tab:fixedParams} }
\begin{tabular}{c@{\qquad} c@{\qquad} c}
\hline
\hline
$F_{\textrm{upper}}=i$ 	& $S_i$ 	&  $f_i$ (MHz)\\
\hline
1			& 1/3 	& 4372.399(85) \\
2			& 35/81 	& 4401.771(37) \\
3			& 28/81 	& 4465.172(53) \\
\hline
\hline
\end{tabular}
\end{table}

\section{Results}\label{sec:results}
 \begin{figure}
 \includegraphics[scale=0.9]{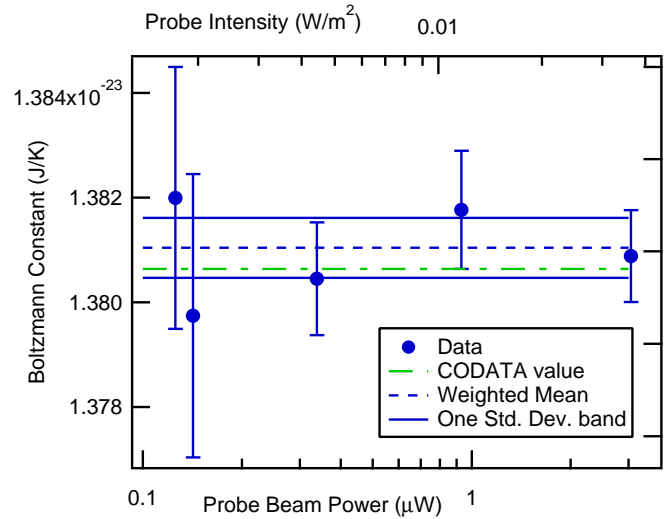}%
 \caption{(Color Online) The value of $k_B$ (dots) extracted from fitting spectra measured at several probe powers. The weighted mean (blue dashed) and weighted one-standard-deviation band (blue solid) are shown. The green, dot-dashed line is the CODATA \cite{moldover} value.\label{graph:results}}%
 \end{figure}
The results from several $k_B$ determinations are shown in Fig. \ref{graph:results} as a function of incident probe power. These data were produced from 24 spectra taken at intensities at least 500-fold below the saturation intensity \cite{steck85} ($I_\textrm{sat}\sim$20 W/m$^2$). The total uncertainties at each power level are indicated by the error bars and were determined from the standard deviation of the data sets taken at the same power level. Our final determination of the Boltzmann constant, based on a weighted mean of the measurements at the various probe powers, is
\[k_B = 1.38104(59)\times 10^{-23} \textrm{J/K}. \]
The uncertainty in this measurement was limited primarily by residual amplitude noise at frequencies higher than the bandwidth of the LCD feedback control loop. This white noise was exacerbated by a lower than expected common-mode rejection of amplitude noise resulting from some asynchronicity in the data acquisition scheme. 

The broadening of the Lorentzian linewidth by magnetic fields contributed a smaller systematic uncertainty of $9.8\times 10^{-5}$, which was summed in quadrature give a total relative uncertainty of $4.1\times 10^{-4}$. All other systematic uncertainties were kept at negligible levels and will be discussed in the following section. However, as the random uncertainties and measurement timing are addressed in future work, these will become increasingly important. Our value of $k_B$ has a relative deviation of $2.7\times 10^{-4}$ from the current CODATA value. To our knowledge, this is the most accurate demonstration of primary thermometry using spectroscopy of \textit{atomic} absorbers.

\section{Estimation of Systematic Uncertainties}\label{sec:systematics}
The greatest source of systematic uncertainty in this experiment was caused by the Earth's magnetic field. This external field lifted the frequency degeneracy of the Zeeman sub-levels in the Rb hyperfine transitions. The field along the cell was measured to be 0.21(1) G and numerical simulations shows that this causes an apparent broadening of the Lorentzian linewidth by 0.5\%, which in turn, contributes a relative uncertainty of $9.8\times 10^{-5}$ in the measured value of $k_B$. Whilst magnetic shields can be used to reduce the field strength by a factor of \mbox{40 000} times, this was not attempted because the effect was smaller than the observed random uncertainties of the experiment.  

The Rb spectra were recorded (non-contiguously) over a one month period. Whilst no drift would be expected from the use of a permanently sealed Rb vapor-pressure reference cell with no buffer gas, checks for long-term stability were performed by periodically re-acquiring spectra at previously investigated probe power levels. We detected no long term drift larger than the short-term reproducibility of the experiment. A long-term drift of the reference laser's wavelength only changes the fitted offset frequency $f_0$ but does not change the Rb spectrum and is therefore inconsequential.  

We also considered the systematic uncertainty caused by holding $\Gamma$ and $S_i$ constant. Fig \ref{graph:chisquare} demonstrates that the minimum of the $\chi^2$ surface lies along $(\Gamma-\Gamma_\textrm{opt})/(\Delta\nu_D-\Delta\nu_{D\textrm{,opt}})\approx 1$, so the propagated relative uncertainty in $\Delta\nu_D$ due to the uncertainty of $\Gamma$ is reduced by a factor $\sim\Delta\nu_D/\Gamma\approx 100$. Since the unperturbed Lorentzian width is known \cite{simsarian1998} with a relative uncertainty of $3\times 10^{-4}$, this contributes a negligible uncertainty in $\Delta\nu_D$. Similarly, there is no uncertainty associated with the transition strengths because they are exact numbers. However, these assumptions are valid only in the limit of zero probe beam power because the measurement process itself can perturb the actual values of $\Gamma$ and $S_i$ away from their theoretical values. For example, excess probe beam power will lead to power-broadening of the underlying Lorentzian linewidth, introducing an intensity dependence in the fitted Doppler width and, thus, in $k_B$. Similarly, any optical pumping between the hyperfine states that produces alignment or polarization of the sample will perturb the effective ratios $S_i/S_j$. We did indeed observe such effects at intensities greater than $\sim$1 W/m$^2$ (at least 30 times larger than intensities reported here), which is still more than an order of magnitude below the saturation intensity \cite{steck85}. 

We have developed a new theory \cite{stace2010} to explain the cause of this early onset of optical saturation behaviour which allows us to confirm that a negligible amount of optical pumping occurred in the experiments reported here. In contrast to previous models \cite{harris2006,Shin2005,siddons} that approximate the atom as a two-level system under uniform illumination, we developed a theory for multi-level atoms probed by a beam with finite spatial extent \cite{stace2010}. This better captures the physical properties and situation of this Rb experiment. We find that the saturation-like behaviour comes from an optical pumping process that evacuated the population in the laser-coupled ground state, and not from a Rabi-flopping mechanism responsible for saturation in a two-level system \cite{siegman,demtroeder1981}.We, therefore, quantify the perturbation away from thermal equilibrium by introducing a figure-of-merit, denoted $\mathscr{F}$, defined as the relative depletion of the laser-coupled ground state, i.e.
\begin{eqnarray}
\mathscr{F}\equiv 1-\frac{\rho_1}{\rho_{1,\textrm{th}}} \label{eqn:FoM}
\end{eqnarray}
where $\rho_{1,\textrm{th}}$ and $\rho_1$ are the thermal and pumped fractional occupation of the ground state. In the absence of optical pumping $(\rho_1=\rho_{1,\textrm{th}})$, $\mathscr{F}=0$; whilst conversely, when the ground state is fully depleted, $(\rho_1=0)$ and $\mathscr{F}=1$. Using the full model of Stace and Luiten \cite{stace2010}, we computed the figure-of-merit to be $\mathscr{F}\approx 0.021$ for the largest powers in this experiment. In Appendix \ref{sec:altFoM}, we show how $\mathscr{F}$ may be calculated approximately using a simple method based on conservation of energy considerations. At $\mathscr{F}=0.021$, the perturbation in $k_B$ due to optical pumping is 4 ppm (see Appendix \ref{sec:perturbation}),  which is well below the current statistical uncertainty. We also showed experimentally that we were sufficiently close to the zero-probe-power limit by demonstrating the absence of any statistically significant slope in the deduced value of $k_B$ as a function of power on Fig. \ref{graph:results}.

\section{Comparison to Molecular DBT}
\begin{table}
\caption{Different experimental regimes accessed by various choices of absorbers. The relative noise reported in the bottom two rows are given for a 1000 s bandwidth.\label{Tab:compare}}
\begin{tabular}{l |c| c| c}
\hline
\hline
			& Rb	& NH$_3$ \cite{daussy2007}	& CO$_2$ \cite{casa2008} \\
\hline
Doppler Width (MHz)	& 308	& 50	& 275\\
Pressure (Pa)	& $10^{-5}$	& 8	& 130\\
Pressure Broadening (MHz)	& 0.02	&1	& 4\\
Optical Pumping, $\mathscr{F}$	& 0.021	& $8\!\!\times\!\!10^{-8}$	& $3\!\!\times\!\!10^{-7}$\\
Probe Power ($\mu$W)	& 0.1	& 0.1	& 50\\
1000 s Shot Noise Floor (ppm) 	& 0.7	& 0.8	& 1.2\\
1000 s Achieved Reproducibility (ppm)	& 4000	& 1000	& 200\\
\hline
\hline
\end{tabular}
\end{table}

In this section, we will highlight the differences between a low-pressure atomic approach and high-pressure molecular DBT experiments by examining reported results, simulated spectra and the results presented in Section \ref{sec:results}. Table \ref{Tab:compare} shows a comparison of some recent DBT experiments using molecules and atoms.

\subsection{Shot-noise limits}
Using the probe powers shown in Table \ref{Tab:compare}, row 5, we simulated shot-noise limited Doppler-broadened spectra consisting of 1000 absorbance samples spanning $f_0\pm 3\Delta\nu_D$. An approximate estimate of the shot-noise limited relative uncertainty of $k_B$ was found by fitting these spectra to our model. The results are surprisingly similar for all experiments (row 6).  The results depend mostly on the total scan time and not details such as the frequency spacings between points. Despite the high CO$_2$ probe power reported in Ref.  \cite{casa2008}, the shot-noise limit is similar to the other experiments because of the low absorption depth in that experiment. The last row of Table \ref{Tab:compare} shows that the approximate actual repeatability calculated from reported results, when scaled to 1000 s time scale, is significantly worse than the shot-noise limit in all cases. Djerroud \textit{et. al.} \cite{djerroud2009} have reported recent improvements to the work of Daussy \textit{et. al.} by achieving a reproducibility of 100 ppm at 1000 s. In our case, we found that background light levels have a significant impact on experimental repeatability, so we intend in future experiments to employ synchronous detection of a modulated carrier. Preliminary results suggest this improves the repeatability to 150 ppm. 

\subsection{Optimization of Probe Beam Power and Geometry}
Current approaches in DBT experiments range in pressures\cite{djerroud2009,casa2008} from $10^{-5}$ to 100 Pa. In the Rb experiment described here, the mean free path is 120 m and the gas dynamics is governed by an effusive flow of atoms, i.e. the system is essentially collisionless, leading to negligible pressure-induced self-broadening \cite{demtroeder1981}. In contrast, the molecular DBT experiments are dominated by collisions and is in the diffusive regime, causing $\sim 0.01\Delta\nu_D$ of pressure-related broadening \cite{daussy2007,lebarbu2006}. Moreover, the physical details of the molecule-molecule collisions are critical in determining the observed lineshape profile. For example, the  degree of elasticity of molecular collisions perturbs the lineshape away from a Voigt profile \cite{wende1981}. This requires higher-order forms such as the Galatry profile and a greater number of adjustable parameters required for the regression of spectra. 

\begin{figure}
 \includegraphics[scale=0.9]{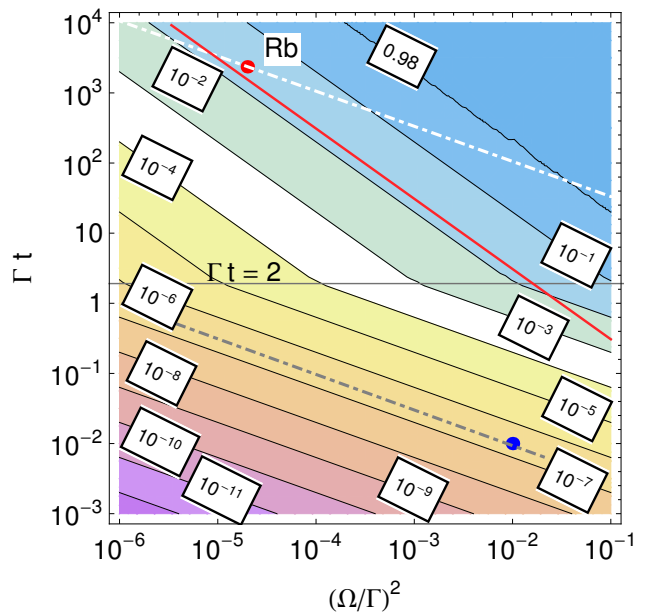}%
 \caption{(Color Online) A contour plot of $\mathscr{F}$ computed from the model of Stace and Luiten\cite{stace2010}, as a function of normalized probe intensity $(\Omega/\Gamma)^2$ and normalised coherent evolution time ($\Gamma t$). The red and blue dots indicate the conditions of this and the molecular DBT experiments, respectively, demonstrating the exploration of vastly different spectroscopic regimes. The dot-dashed lines are contours of constant \textit{input} power. The solid red line indicates the experimental conditions that would lead to a 1 ppm perturbation in $k_B$. \label{fig:pumping_map}}%
 \end{figure}

In contrast to the collisional effects of the molecular DBT systems the critical linewidth perturbation for vapours with strong light interaction is associated with optical pumping. This is clearly demonstrated in row 4 of Table \ref{Tab:compare}. To illustrate the very different operational spectroscopic regimes interrogated in this work and by Daussy \textit{et. al.} \cite{daussy2007,djerroud2009} and Casa \textit{et. al.} \cite{casa2008,castrillo2009}, Fig. \ref{fig:pumping_map} shows a contour plot of $\mathscr{F}$ as a function of the non-dimensional quantities $\Gamma t$ and $(\Omega/\Gamma)^2$. The vertical axis corresponds to the ratio of the coherent interaction time, $t$, to its lifetime ($1/\Gamma$): for an effusive vapor like Rb, $t$ is the average beam transit time while for a diffusive gas, like those used by the molecular experiments, $t$ is the mean time between collisions. The horizontal axis is proportional to the beam intensity, and is expressed in terms of the Rabi frequency, $\Omega$, and natural decay rate, $\Gamma$. The conventional intensity can be related to these parameters by the expression $I = 2 (\Omega/\Gamma)^2 I_\textrm{sat}$, where $I_\textrm{sat}$ is the conventional 2-level saturation intensity\cite{steck85}. The $\mathscr{F}$ contours, which were calculated with the model of Stace and Luiten\cite{stace2010}, can be closely approximated by the simpler expressions:
%The vertical axis is in units of a non-dimensional coherent interaction time $\Gamma t$ where $t$ is the average beam transit time for the case of an effusive gas (like Rb) or the mean time between collisions in the case of a diffusive gas (like the molecular experiments). 
%The estimate for $\mathscr{F}$ is based on a simplistic model of a three level atom in a $\Lambda$ configuration, subject to a driving field with Rabi frequency $\Omega$ resonant with one of the transitions for a length of time $t$, and relaxation at a rate $\Gamma/2$ from the excited state to each of the two ground states. It is simple to show that  
\begin{equation}\label{eqn:approxF}
\mathscr{F}\approx\left\{ 
\begin{array}{cc}(\Omega t/2)^2 & \textrm{if } \Gamma t\gg2 \\ 
\Omega^2 t/(2\Gamma) & \textrm{if } \Gamma t\ll2
\end{array}\right. .
\end{equation}
%and it is this function whose contours are plotted in Fig. \ref{fig:pumping_map}.  This simple model gives a reasonable estimate of $\mathscr{F}$, which can be refined using more sophisticated models capable of predicting $\mathscr{F}$ quantitatively \cite{stace2010}. 
Thus, contours of constant $\mathscr{F}$ have a slope of $-1$ decade/decade and $-1/2$ decade/decade when $\Gamma t\gg 2$ and $\Gamma t\ll 2$, respectively. Eqn. \ref{eqn:approxF} was derived using a model of a three level atom with two ground states and an excited state. The scales of Fig. \ref{fig:pumping_map} obscure the details of the more complex case $\Gamma t \approx 2$ , which are unimportant to this wide range overview.

In Fig. \ref{fig:pumping_map}, the solid red line indicates the parameters where there would be a 1 ppm perturbation to the determination of $k_B$ (derived from the analysis in Appendix \ref{sec:perturbation}). For collisionless gases, $t$ is proportional to the beam radius $r$ and $(\Omega/\Gamma)^2\propto r^{-2}$. Therefore, changing the probe beam diameter for a fixed input power lead to lines with a slope of $-1/2$ decade/decade. Two such contours (dot-dashed) are drawn through the location of the Rb experiment reported here as well as for the molecular experiments. Contours of constant \textit{absorbed} power follow lines of constant $\mathscr{F}$ since the absorption coefficient is approximately proportional to the population perturbation in the ground state, i.e.
\begin{eqnarray}
V(f_i, \Delta\nu_D) \propto \rho_1 = \rho_{1,\textrm{th}}(1-\mathscr{F}).\label{eqn:absorbed_power}
\end{eqnarray} 

It is clear from Fig. \ref{fig:pumping_map} that there are two regimes of coherent interaction time, probe intensity and optical pumping which are delineated by the line $\Gamma t = 2$.  For $\Gamma t \ll 2$, the contours of $\mathscr{F}$ and those for fixed input power are parallel, demonstrating that the measured results will be independent of the probe beam geometry. However, this does not hold when $\Gamma t \gg 2$, indicating that the result will depend on probe beam shape. In this situation, Fig. \ref{fig:pumping_map} guides the experimenter into an optimisation of  probe beam radius and intensity for a given perturbation to $k_B$. By increasing the beam radius, it is possible to increase the \emph{total input power} whilst remaining on the red line. This is advantageous as it will lead to an improved SNR of the measurement for a given linewidth perturbation. The limit to this procedure is set by the size of the gas cell. For molecular experiments where the coherent interaction time is limited by the mean time between collisions (and not the beam transit time), an increase in probe power will require a corresponding increase in pressure to preserve the amount of optical pumping. Of course, this compensation may be forgone given the much smaller magnitude of optical pumping in this regime. However, the small optical cross sections of the molecular transitions which reduce the impact of optical pumping effects give rise to other issues associated with SNR and pressure-broadening. 

\subsection{Absorption Depth and SNR}
Whilst Fig. \ref{fig:pumping_map} demonstrates the impact of experimentally controllable variables such as the coherent interaction time (which can be adjusted by changing the beam radius in an effusive gas system, and by the pressure in a diffusive gas) and the probe beam intensity, there are other considerations pertaining to the atomic/molecular properties that independently affect SNR. One such consideration is the optical depth. Numerical simulations show that in the presence of white amplitude noise (such as detector noise) there is an optimum optical depth, $\alpha L\approx 3$ at which the variance in the Doppler width obtained from repeated simulations is minimised. In terms of fundamental parameters, the optical depth can be expressed as\cite{siegman}
\begin{eqnarray}
\alpha L =\sigma_0\rho_0 L\propto\Gamma\lambda^2 P L, 
\end{eqnarray}
where $\sigma_0$, $\rho_0$, $P$ and $L$ are the optical cross-section, number density, pressure of the absorber and optical path length, respectively. We can more clearly see the trade-offs between the parameters that affect the optical depth by writing an ``equation of state'' for the optical depth using the conditions of the Rb experiment as the reference values,
\begin{eqnarray}
\left(\frac{\alpha L}{0.7} \right) &&\sim \nonumber\\ \left(\frac{(\lambda/\textrm{nm})^2}{780^2}\right)&&\left(\frac{\Gamma/\textrm{s}^{-1}}{4\times 10^7}\right)\left(\frac{P/\textrm{Pa}}{3.5\times 10^{-5}}\right)\left(\frac{L/\textrm{m}}{10^{-1}}\right). 
\end{eqnarray}
To achieve comparable SNR to atomic absorbers in which $\Gamma$ is large, molecular-based approaches must compensate by increasing the product of $P$ and $L$ because of the much longer lifetimes. Increasing either of these parameters can lead to the potential for unwanted inaccuracies in the experiment relating to pressure broadening or thermal gradients in the apparatus.

%Whilst it was possible in the Rb experiment to explore a region of smaller $\mathscr{F}$ by reducing the input power (e.g. moving vertically downwards in Fig. \ref{fig:pumping_map} by decreasing the beam radius for a set value of intensity), it was not feasible to reach a value equivalent to those achieved by the molecular experiments ($\sim 10^{-7}$) because of the rapid decrease in SNR. Similarly, the molecular experiments cannot access the strong pumping regime since it requires $\sim 100$-fold increase in the beam radius to do so. Perhaps operating at the geometric mean of the extremes defined by the existing experiments might provide simultaneous insensitivity to pressure related and optical pumping effects. For example, there are a series of suitable transitions between 311 nm and 323 nm in Rb. In particular, the 315.75 nm transition \cite{kratz1949} of Rb has $\Gamma=3.4\times 10^4$, which is approximately 1000 times less absorbing than the lines used in this study.

\subsection{Windows of Operation}

 \begin{figure}
 \includegraphics[scale=1]{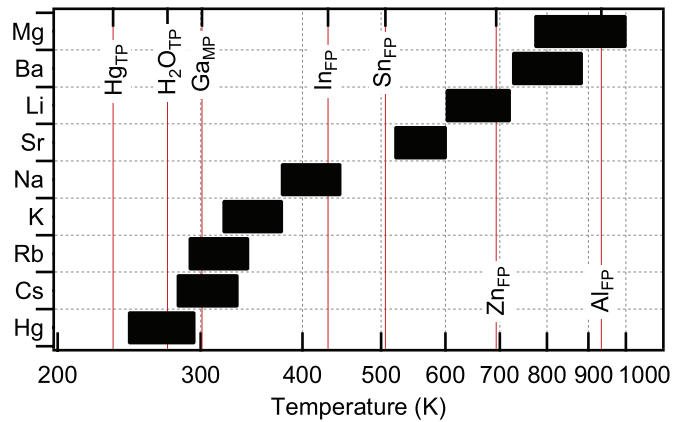}%
 \caption{(Color online) Approximate temperature windows of operation for various vapor-pressure cells, of length restricted to 10 cm. The species in each cell is shown on the vertical axis. The temperatures of selected ITS-90 fixed points are indicated by the solid vertical red lines. The subscripts TP, MP and FP denote Triple Point, Melting Point and Freezing Point, respectively.  \label{fig:windows}}%
 \end{figure}

In this experiment, we used a commercially available Rb vapor cell of a standard length (10 cm), which gave an absorption depth of $\sim 50\%$ at room temperature. However, operating along the metal's sublimation curve means that the equilibrium vapor density increases exponentially with temperature, and thus the absorption depth is much more sensitive to temperature for atomic systems than in the molecular experiments. The range of temperatures over which we can use Rb as a thermometric substance is comparatively smaller than for molecules. The lower temperature bound is provided by the diminishing absorption depth resulting from the exponential decrease in vapor pressure. Conversely, the upper temperature bound is set when the vapor density becomes so large that the probe beam is completely absorbed over a small range of frequency detuning. Fig. \ref{fig:windows} shows approximate temperature windows over which a selection of metal gases at their vapor pressure might be able to operate \cite{steck85,CRC,pilling1921,rudberg1935,gilbreath1965,huber2006,xu2003,asano1978}. At the upper and lower temperature bounds of each window, the irreproducibility of the fitted Doppler width in repeated simulations is double that for the optimal optical depth (i.e. $\alpha L\approx 3$). This condition restricts the optical depth to a range $0.5 < \alpha L < 90$. We have restricted attention to 10 cm long cells for convenient thermal control. Some fixed points of the ITS-90 \cite{fischer2005} scale are also shown, demonstrating that metallic vapor-pressure DBT can be used to verify the thermodynamic temperature at those fixed points. 

For a particular choice of absorber, it is also possible to fill the reference cell using an external atomic reservoir at a lower temperature, resulting in a lower vapor number density. Once the vapor cell is sealed, no further increase in number density is possible, thereby restricting optical depth to an optimum level when at high temperatures. This would permit the exploration of temperatures and pressures away from the sublimation phase boundary. 

\section{Conclusions}
Doppler broadening thermometry provides a significantly different approach to the determination of the Boltzmann constant in preparation for the re-definition of the kelvin. Whilst previous DBT experiments have studied molecular absorbers at up to 10 Pa in a diffusive regime, we have used Rb at $3\times 10^{-5}$ Pa where the gas dynamics are effusive. Our approach avoids problems such as pressure broadening and poor signal-to-noise. Furthermore, the Rb cell is compact in size, enabling more convenient temperature control. We used this system to determine $k_B$ with a relative uncertainty of $4\times 10^{-4}$. The present experiment was limited by amplitude noise in the probe beam, which in future work will be overcome by using a control loop with larger bandwidth. We estimate that all current DBT experiments are far from the shot noise limit and that the use of synchronous detection methods could address this situation.

We also compared sources of systematic uncertainty for DBT experiments in the diffusive regime of molecular DBT and the effusive regime of atomic DBT. For atomic vapors such as Rb, the shift in the determined value of $k_B$ due to the Earth's magnetic field is approximately 100 ppm. Optical pumping effects are determined by the combination of beam transit time, upper-state lifetime and probe intensity. In this work, the equilibrium ground-state population was perturbed by 0.2\% which shifted the measured $k_B$ value by 4 ppm. The small cross-sections of the transitions used in molecular DBT experiments means that the optical pumping effects are relatively small (perturbation to equilibrium ground-state population of order $10^{-7}$). However, these small cross-sections mean that an equivalent signal-to-noise in molecular experiments requires long path lengths and/or higher pressures, both of which have associated systematic effects on measurements of $k_B$. The method developed here for quantifying optical pumping effects can be used to search for candidate absorbers optimally suited to DBT experiments in either the diffusive or effusive regimes.

\appendix
\section{Approximate method for calculating $\mathscr{F}$}\label{sec:altFoM}
Whilst the approach of Stace and Luiten \cite{stace2010} captures the details of the gas-dynamical interaction between atoms and the beam volume, which can be used to estimate the population perturbation using no phenomenological parameters, it relies on a knowledge of the dipole matrix element in order to calculate the Rabi frequency. It is useful to compare this against a simpler, independent analytic calculation. We provide such a calculation here, based on energy conservation and using an \textit{a posteriori} knowledge of the optical depth. 

Assuming that the input intensity is sufficiently weak so that an atom never absorbs more than one photon during a single transit of the beam, the rate of atoms scattered out of the laser-coupled ground state is approximately the same as the rate of scattered photons. This is given by
\begin{eqnarray}
 R = a \frac{I_0}{E_\gamma}(1-I/I_0),
\end{eqnarray} 
where $a=\pi r^2$ is the cross-sectional area of a beam with radius $r$, $E_\gamma$ is the photon energy and $I$ and $I_0$ are the \textit{on-resonance} output and input intensities. 

Using kinetic theory \cite{reif1965}, the flux of atoms crossing the beam surface (of area $A=2\pi r l$, where $l$ is the cell length) is
\begin{eqnarray}
 \Phi A = \frac{1}{6}\bar{v} \rho_0 \frac{\Gamma}{\Delta\nu_D} A, 
\end{eqnarray}
where $\bar{v}=\sqrt{2kT/m}$ is the mean atomic speed and $\rho_0$ is the total number density. A factor of $\Gamma/\Delta\nu_D$ is used to count only those atoms which are \textit{on-resonance}. 

The value of $\mathscr{F}$ (Eqn. \ref{eqn:FoM} from Section \ref{sec:systematics}) is then the ratio of the rate of scattering to the rate of new atoms impinging on the beam, i.e.
\begin{eqnarray}
 \mathscr{F} &=& R/(\Phi A) \nonumber\\
	&=& \frac{3r}{\bar{v}\rho_0 l}\frac{\Delta\nu_D}{\Gamma}\left(\frac{I_0}{E_\gamma}\right)\left(1-\frac{I}{I_0}\right)\\
	&\approx& 0.022\qquad\textrm{for this experiment.} \nonumber
\end{eqnarray}
This approximation agrees with the more detailed analysis, differing only by $10^{-3}$.

 \begin{figure}
 \includegraphics[scale=1]{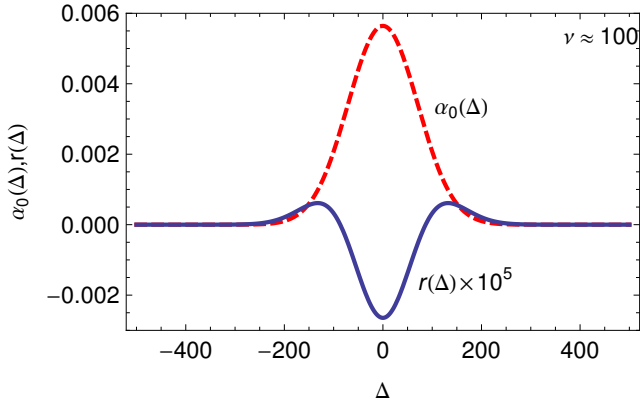}%
 \caption{(Color Online) An example of the unperturbed absorption coefficient ($\alpha_0(\Delta)$) and the additive perturbation term ($r(\Delta)$), plotted for $\nu \approx 30 \textrm{MHz}/3 \textrm{MHz} = 100$. The perturbation term is vertically expanded by $10^{5}$ for clarity. Note that $\nu$ in this Appendix has a different meaning to that used in the preceding sections of this paper.\label{fig:perturbation}}%
 \end{figure}

\section{Linewidth Perturbation}\label{sec:perturbation}
Using the approach and notation of Stace and Luiten \cite{stace2010}, the general expression for the absorption coefficient (Eqn. (22) of Ref.  \cite{stace2010}) can be expanded (Appendix A1) in the case of small optical pumping. The expression for a perturbed absorption coefficient lineshape is
\begin{eqnarray}
\alpha&\approx&\frac{\kappa c P_1 z}{2\omega}\left(\left\{\frac{P_3''-P_1''}{P_1}\frac{\sqrt{\nu^2+1/4}}{8\nu}-\pi\right\}V(\Delta,\nu)\right.\nonumber\\
& &\left.+\frac{P_3''-P_1''}{P_1}\frac{4(\Delta/\nu)^2-3}{64\sqrt{\pi}\nu^3}\exp\left(-\frac{\Delta^2}{\nu^2+1/4}\right)\right) \nonumber\\
&\approx& \alpha_0(\Delta)+\epsilon r(\Delta),
\end{eqnarray}
where $\alpha_0(\Delta)$ is the unperturbed lineshape, $r(\Delta)$ is defined as the exponential term in the parentheses and $\Delta$ is the detuning in units of the Gaussian width. Fig. (\ref{fig:perturbation}) shows a plot of the unperturbed lineshape along with the perturbation term $r(\Delta)$. We performed an expansion in the small parameter $\epsilon\equiv (P_3''-P_1'')/P_1$, where $P_i''\equiv\partial^2P_i/\partial\Delta^2$ the derivative is with respect to the detuning $\Delta$. The meanings of the remaining variables in this section are as defined in Ref.  \cite{stace2010}. Written in this form, it is clear that the absorption coefficient in the presence of some optical pumping is a Voigt function, denoted $V(\Delta,\nu)$ with an additional perturbation term.  

Consider the expansion of a characteristic width for the perturbed absorption coefficient,
\begin{eqnarray}
\Delta_\textrm{1/e,pert}(\epsilon)\approx\Delta_\textrm{1/e,pert}(0) + \left.\frac{d \Delta_\textrm{1/e,pert}}{d\epsilon}\right|_{\epsilon=0}\delta\epsilon, \label{eqn:linewidthpert}
\end{eqnarray}
such that $\alpha(\Delta_\textrm{1/e,pert})/\alpha(0)\equiv 1/e$. Differentiating this definition and using $\Delta_\textrm{1/e,pert}(0)= \Delta_{1/e}$, we obtain the perturbation to the linewidth
\begin{eqnarray}
\left.\frac{d \Delta_\textrm{1/e,pert}}{d\epsilon}\right|_{\epsilon=0} &=& \frac{\alpha_0(0)[r(0)/e - r(\Delta_{1/e})]}{[\alpha_0(0)+\epsilon r(0)][\alpha_0'(\Delta_{1/e})+\epsilon r'(\Delta_{1/e})]} \nonumber\\
&\approx& \frac{a_0(0)[r(0)/e-r(\Delta_{1/e})]}{a_0 a_0'(\Delta_{1/e})}\nonumber\\
&\approx& -\frac{1/e}{16\sqrt{\pi}\nu^3}\frac{1}{\alpha_0'(\Delta_{1/e})}\nonumber\\
&\approx& -\frac{1}{32\sqrt{\pi}\nu^2} .
\end{eqnarray}

To evaluate $\delta\epsilon$, we have used the low probe power approximation of Stace and Luiten ($P_3\approx 0 <<P_1$) and $\Delta_{1/e}\approx\nu\equiv \Delta\nu_D/\Gamma\approx 100 >> 1$. We also used the approximate form for the steady-state population density $P_1$ that ignores atomic coherence. This quantity is maximal on resonance (i.e. when $\Delta=0$): 
\begin{eqnarray*}
\textrm{Max}[P_1''] = P_1''(0) &=& \left.\frac{d^2}{d\Delta^2}\left(\frac{1}{2}(1-\exp(-\frac{\Omega^{*2}t^*}{1+4\Delta^{*2}}))\right)\right|_{\Delta=0}\\
&=& -4t^*\Omega^{*2} \exp(-t^*\Omega^{*2})\\
&<& -4t^*\Omega^{*2}\\
&\approx& -0.4\qquad\textrm{for this experiment.}
\end{eqnarray*}

Substitution of the relevant quantities into Eqn. (\ref{eqn:linewidthpert}) gives 2 ppm perturbation to the linewidth and, therefore, 4 ppm in the determination of $k_B$ for this experiment.

% If in two-column mode, this environment will change to single-column format so that long equations can be displayed. 
% Use only when necessary.
%\begin{widetext}
%$$\mbox{put long equation here}$$
%\end{widetext}

% Figures should be put into the text as floats. 
% Use the graphics or graphicx packages (distributed with LaTeX2e).
% See the LaTeX Graphics Companion by Michel Goosens, Sebastian Rahtz, and Frank Mittelbach for examples. 
%
% Here is an example of the general form of a figure:
% Fill in the caption in the braces of the \caption{} command. 
% Put the label that you will use with \ref{} command in the braces of the \label{} command.
%

% \begin{figure}
% \includegraphics{testimage}%
% \caption{blah\label{graph:staircase}}%
% \end{figure}

% Tables may be be put in the text as floats.
% Here is an example of the general form of a table:
% Fill in the caption in the braces of the \caption{} command. Put the label
% that you will use with \ref{} command in the braces of the \label{} command.
% Insert the column specifiers (l, r, c, d, etc.) in the empty braces of the
% \begin{tabular}{} command.
%

% If you have acknowledgments, this puts in the proper section head.
%\begin{acknowledgments}
% Put your acknowledgments here.
%\end{acknowledgments}

%% Create the reference section using BibTeX:
%\bibliography{../bib2}
%\bibliographystyle{unsrt}

\end{document}